\documentclass[%
reprint,
superscriptaddress,
amsmath,amssymb,
aps,longbibliography,
]{revtex4-2}

\usepackage{graphicx}%
\usepackage{verbatim}
\usepackage[version=4]{mhchem}
\usepackage{xcolor}
\usepackage{physics}
\usepackage[normalem]{ulem}
\usepackage{pdfpages}
\usepackage{pgffor}
\usepackage{booktabs}

\usepackage{listings}
\usepackage{xcolor}
\usepackage{tcolorbox}

\usepackage{listings}

\makeatletter
\AtBeginDocument{\let\LS@rot\@undefined}
\makeatother

\begin{document}

\title{Massive Atomic Diversity: a compact universal dataset for atomistic machine learning}

\author{Arslan Mazitov}
\email{arslan.mazitov@epfl.ch}
\affiliation{Laboratory of Computational Science and Modeling, Institut des Mat\'eriaux, \'Ecole Polytechnique F\'ed\'erale de Lausanne, 1015 Lausanne, Switzerland}

\author{Sofiia Chorna}
\affiliation{Laboratory of Computational Science and Modeling, Institut des Mat\'eriaux, \'Ecole Polytechnique F\'ed\'erale de Lausanne, 1015 Lausanne, Switzerland}

\author{Guillaume Fraux}
\affiliation{Laboratory of Computational Science and Modeling, Institut des Mat\'eriaux, \'Ecole Polytechnique F\'ed\'erale de Lausanne, 1015 Lausanne, Switzerland}

\author{Marnik Bercx}
\affiliation{PSI Center for Scientific Computing, Theory and Data, 5232 Villigen PSI, Switzerland}
\affiliation{National Centre for Computational Design and Discovery of Novel Materials (MARVEL), 5232 Villigen PSI, Switzerland}

\author{Giovanni Pizzi}
\affiliation{PSI Center for Scientific Computing, Theory and Data, 5232 Villigen PSI, Switzerland}
\affiliation{National Centre for Computational Design and Discovery of Novel Materials (MARVEL), 5232 Villigen PSI, Switzerland}

\author{Sandip De}
\affiliation{BASF SE, Carl-Bosch-Stra{\ss}e 38, 67056 Ludwigshafen, Germany}

\author{Michele Ceriotti}
\email{michele.ceriotti@epfl.ch}
\affiliation{Laboratory of Computational Science and Modeling, Institut des Mat\'eriaux, \'Ecole Polytechnique F\'ed\'erale de Lausanne, 1015 Lausanne, Switzerland}

\newcommand{\CSM}[1]{{\color{red}#1}}

\date{\today}%

\begin{abstract}
The development of machine-learning models for atomic-scale simulations has benefitted tremendously from the large databases of materials and molecular properties computed in the past two decades using electronic-structure calculations. 
More recently, these databases have made it possible to train ``universal'' models that aim at making accurate predictions for arbitrary atomic geometries and compositions.
The construction of many of these databases was however in itself aimed at materials discovery, and therefore targeted primarily to sample stable, or at least plausible, structures and to make the most accurate predictions for each compound -- e.g. adjusting the calculation details to the material at hand. 
Here we introduce a dataset designed specifically to train machine learning models that can provide reasonable predictions for arbitrary structures, and that therefore follows a different philosophy. 
Starting from relatively small sets of stable structures, the dataset is built to contain ``massive atomic diversity'' (MAD) by aggressively distorting these configurations, with near-complete disregard for the stability of the resulting configurations. 
The electronic structure details, on the other hand, are chosen to maximize consistency rather than to obtain the most accurate prediction for a given structure, or to minimize computational effort. 
The MAD dataset we present here, despite containing fewer than 100k structures, has already been shown to enable training universal interatomic potentials that are competitive with models trained on traditional datasets with two to three orders of magnitude more structures. 
We describe in detail the philosophy and details of the construction of the MAD dataset. We also introduce a low-dimensional structural latent space that allows us to compare it with other popular datasets, and that can also be used as a general-purpose materials cartography tool. 
\end{abstract}

\maketitle

\section{Background \& Summary}
The introduction of large-scale, open-access materials databases has significantly accelerated computational materials science and discovery \cite{jain+13aplm}. 
They offer vast repositories of atomic structures and computed or experimentally measured properties of organic and inorganic compounds, facilitating high-throughput screening for many materials-discovery applications.
Among these, the databases of electronic structure calculations serve as a particularly important source of data for atomistic modeling, providing a robust and consistent way of exploring structure-property relations for a wide range of materials, including those that have never been experimentally realized \cite{MPtrj, Materials_Cloud, NOMAD, JARVIS, MD22, OC22, Schmidt2023, Wang2023, OMAT24}. 
Despite these advancements, existing datasets primarily focus on structures at, or near to, local minima and saddle points of the potential energy surface (PES), limiting their applicability in atomistic simulations that often require exploration of mid- and high-energy configurations. 
This is particularly important for interatomic potentials — approximations of the PES — which require accurate descriptions of both low- and high-energy states to ensure robustness across a wide range of thermodynamic conditions.
Another source of error stems from the presence of inconsistencies in computational settings between different datasets, and between different structures within the dataset. 
For example, some compositions may be treated with different electronic-structure details to tackle known shortcomings of density-functional theory, which however means that different portions of chemical space are associated with different PES. 
Furthermore, most of the existing datasets are focused on either organic or inorganic materials -- which is well motivated by the fact that these classes of materials often require different electronic-structure details and different energy scales, but restricts the development of universal interatomic potentials capable of handling hybrid systems of various nature and chemical compositions. 

\begin{table*}[tbp]
\centering
\begin{tabular}{@{}lp{0.5\textwidth}rr@{}}
\toprule
\textbf{Subset name} & \textbf{Description} & \textbf{\# structures} & \textbf{\# atoms} \\
\midrule
MC3D & Bulk crystals from the Materials Cloud 3D crystals database~\cite{Huber2022MC3D} & 33596 & 738484 \\
MC3D-rattled & Rattled analogs of the original MC3D crystals, with Gaussian noise added to all atomic positions & 30044 & 599675 \\
MC3D-random & Artificial structures from MC3D with randomized atomic species from 85 elements & 2800 & 25095 \\
MC3D-surface & Surface slabs generated from MC3D by cleaving along random low-index crystallographic planes & 5589 & 205185 \\
MC3D-cluster & Nanoclusters (2-8 atoms) cut from MC3D and MC3D-rattled environments & 9071 & 44829 \\
MC2D & Two-dimensional crystals from the Materials Cloud 2D database~\cite{moun+18nn,Campi2023MC2D} & 2676 & 43225 \\
SHIFTML-molcrys & Curated SHIFTML molecular crystals from the Cambridge Structural Database~\cite{Cordova2022, groom_cambridge_2016} & 8578 & 852044 \\
SHIFTML-molfrags & Neutral molecular fragments from the SHIFTML dataset~\cite{Cersonsky2023} & 3241 & 72120 \\
\bottomrule
\end{tabular}
\caption{Description of structural subsets used that constitute the MAD dataset.
}
\label{tab:structure-subsets}
\end{table*}

To address these challenges, we introduce the Massive Atomic Diversity (MAD) dataset, designed to encompass a broad spectrum of atomic configurations, including both organic and inorganic systems, while being restricted to a small number of structures -- which facilitates property estimation with converged settings, and reduces the cost of training new models based on it. 
By applying systematic perturbations to stable structures and maintaining consistent computational parameters, we aim to provide a coherent structure-energy mapping suitable for training robust, general-purpose machine learning interatomic potentials suitable for complex atomistic simulation workflows. In the following sections, we detail the construction methodology of the MAD dataset, analyze its diversity and consistency, and demonstrate its efficacy in training machine learning models that perform competitively with those trained on significantly larger traditional datasets.

\begin{figure}[tbp]
    \centering
      \includegraphics[width=\linewidth]{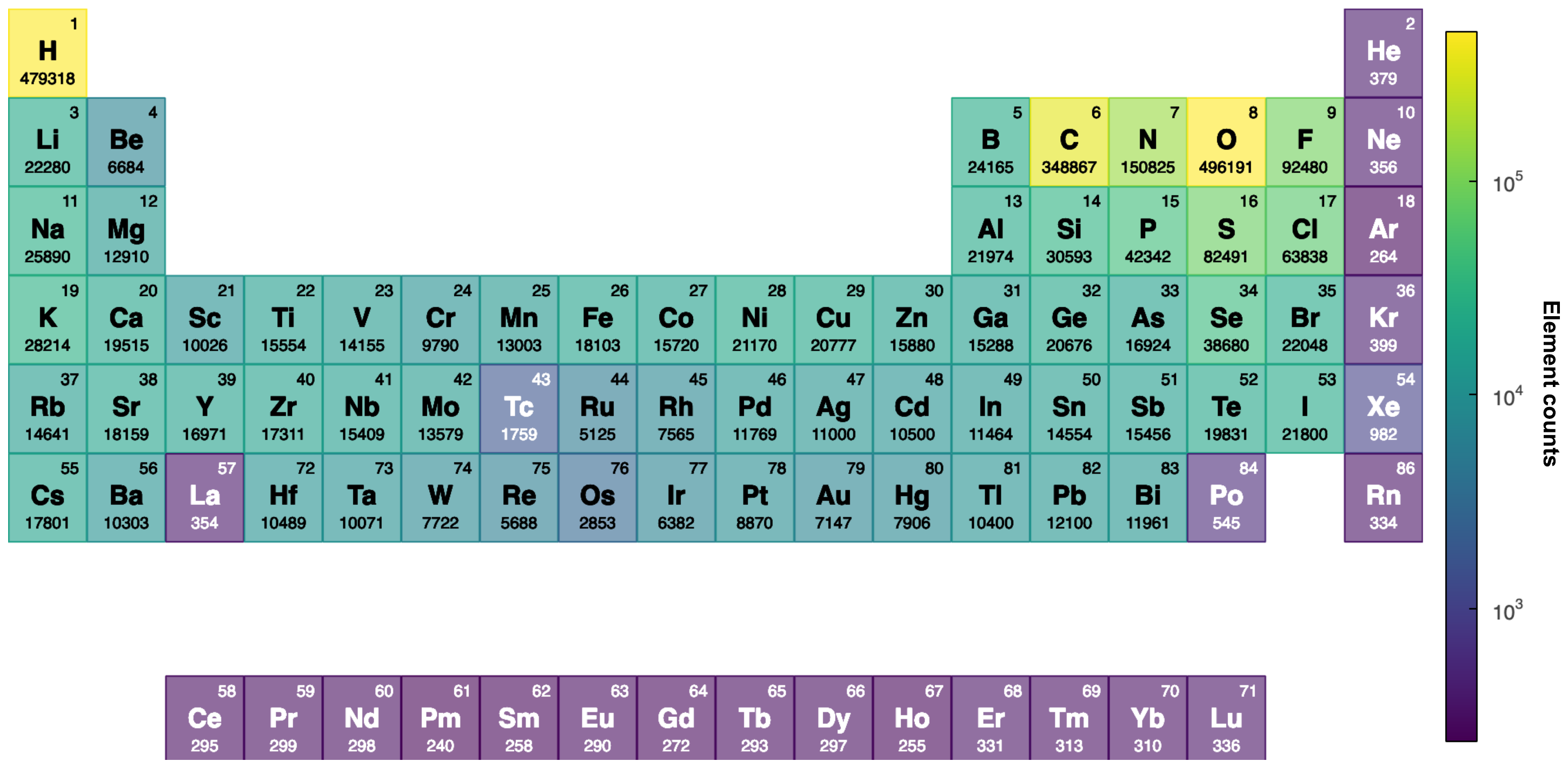}
    \caption{Periodic table indicating the statistical representation of the elements present in the MAD dataset.
    }
    \label{fig:mad-elements-occurences}
\end{figure}

\section{The MAD dataset}

In contrast to most existing datasets for atomistic machine learning, which usually contain stable --- or judiciously distorted --- configurations of materials, focusing primarily on either inorganic or organic domains, the MAD dataset is based on a different philosophy. It draws inspiration from the \emph{mindless dataset} proposed by Korth and Grimme to benchmark quantum chemistry methods~\cite{kort-grim09jctc}. 
First, it aims to extend the limits of universality by incorporating both organic and inorganic materials, thus allowing the creation of models capable of performing atomistic simulations in both domains.
Second, it systematically extends the coverage of the configuration space by adding relaxed structures, their rattled counterparts, structures with randomized composition, clusters, molecules and surfaces, enabling complex simulation protocols in a broad range of thermodynamic conditions, including out of equilibrium conditions (see Table~\ref{tab:structure-subsets} for an overview of the different subsets of structures included in MAD). 
Third, it uses a consistent level of theory across all \textit{ab initio} calculations to ensure a coherent structure-energy mapping for the included structures.  
This unfortunately means the MAD dataset neglects the description of a few important physical effects, such as magnetism, electron correlations, and dispersion, which can be important for certain types of materials yet cannot be applied consistently across the MAD dataset. 
Section~\ref{sec:dft_details} gives more details on the first-principles calculations. 
Last but not least, while maintaining a reasonable descriptive power, the MAD dataset is designed to be lightweight, consisting of fewer than 100,000 structures, thus significantly reducing the total amount of computational resources required for training and making it accessible to a wider community. 

As outlined in Ref.~\citenum{Mazitov2025}, the MAD dataset contains 95595 structures, containing 85 elements in total (with atomic numbers ranging from 1 to 86, excluding Astatine).
The statistical representation of the elements occurrences across the dataset is presented in Figure~\ref{fig:mad-elements-occurences}. Despite being relatively lightweight, MAD provide a good coverage of the main-block elements, with an over-representation of first-period elements, that are abundant in the ``organic'' subsets of MAD. 
Additionally, MAD naturally under represents noble gases due to their low reactivity and limited occurrence in nature, and lanthanides due to technical reasons related to the poor robustness of the underlying DFT calculations -- which means that the reference data is likely to be of low-quality, and therefore of little practical use beside low-accuracy preliminary explorations. More details on this latter issue are provided in Section~\ref{sec:dft_details}.

\begin{figure}[tbp]
    \centering
    \includegraphics[width=1.0\linewidth]{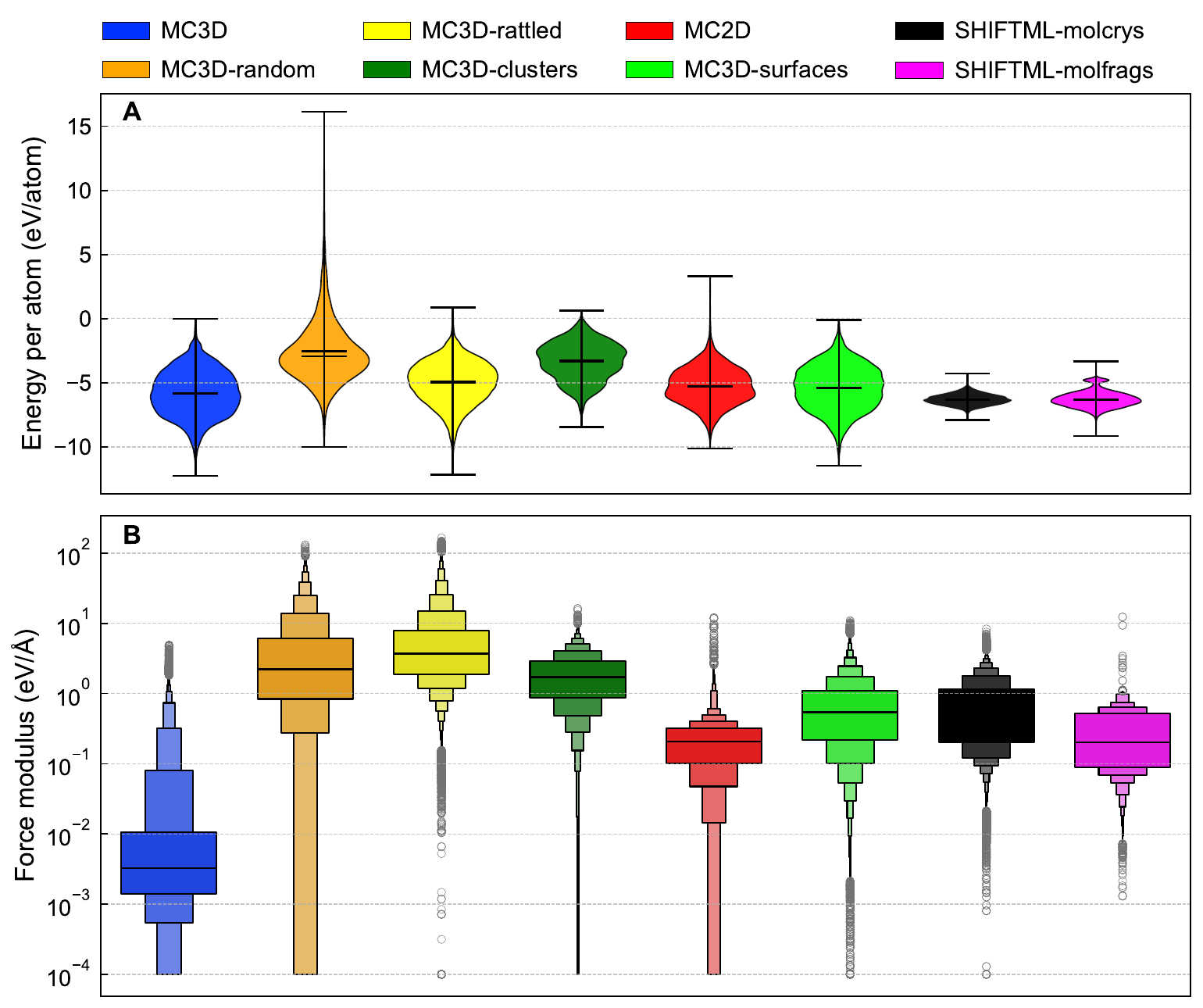}
    \caption{Histograms of energy per atom (top) and force modulus (bottom) within the different subsets of the MAD dataset.    
    \label{fig:data-histograms}
    }
\end{figure}
\subsection{Dataset characterization}

We characterize further the MAD dataset in terms of the distribution of energy and force values within each of the subsets. As shown in Fig.~\ref{fig:data-histograms}, these vary quite wildly: whereas the MC3D, MC2D, SHIFTML datasets are based on stable, or low-temperature MD configurations, and have small interatomic forces, the MC3D-derived subsets, that are built introducing large distortions in the chemical and structural parameters, cover a much larger energy range, which helps obtaining models that are capable of handling highly-distorted, unexpected configurations, and increase the range of structures for which trained model can extrapolate reasonably.

\begin{figure}
    \centering
    \includegraphics[width=1.0\linewidth]{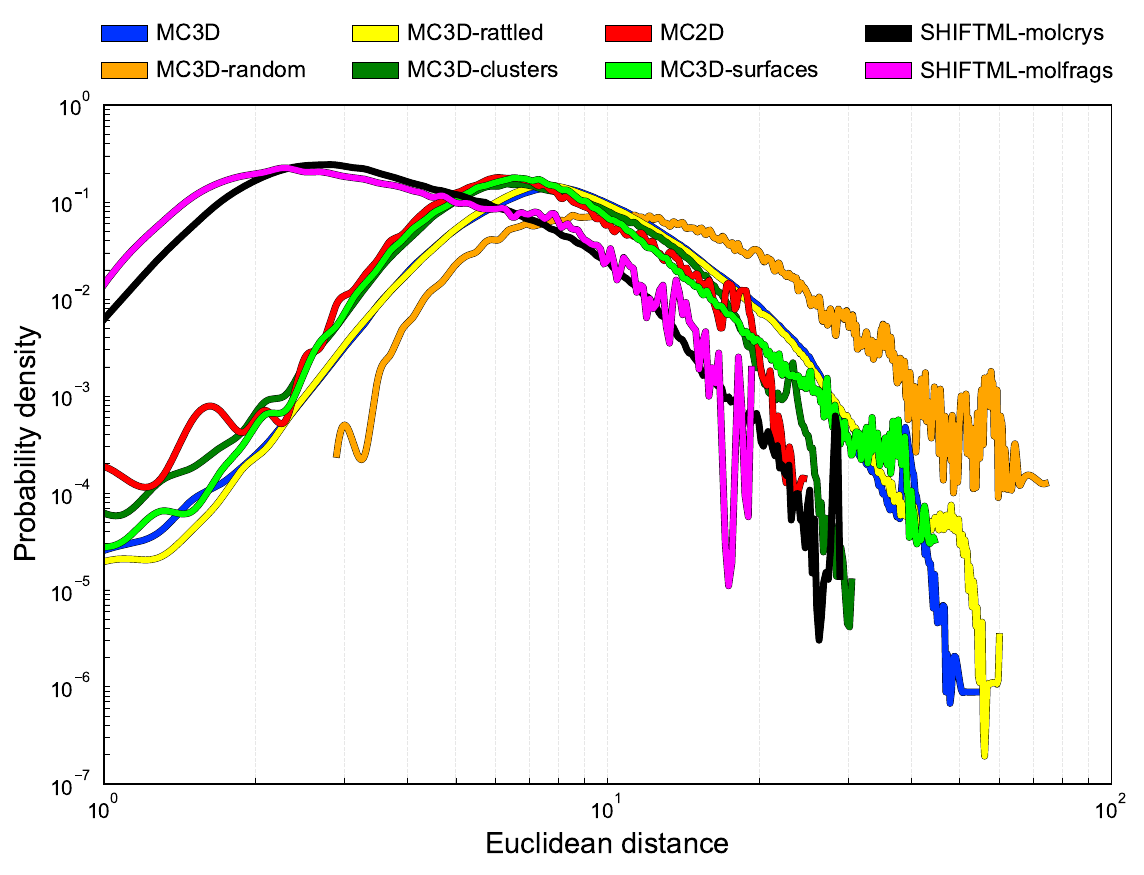}
    \caption{Histogram of the Euclidean distances between structures in the different subsets, computed in the space of high-dimensional PET-MAD features.
    }
    \label{fig:dist-histograms}
\end{figure}

\section{A map of the MAD chemical space}

To substantiate our claim that MAD covers a broader portion of chemical space than existing datasets, and to lay the foundations for a framework to characterize future extensions, and more broadly to represent materials datasets, we proceed to define a low-dimensional representation of the space covered by MAD.
To this end, we need a high-dimensional representation of the structures, and a strategy to reduce the feature space to a dimensionality (2D or 3D) that can be visualized conveniently. 

\begin{figure}[tpb]
    \centering
    \includegraphics[width=\linewidth]{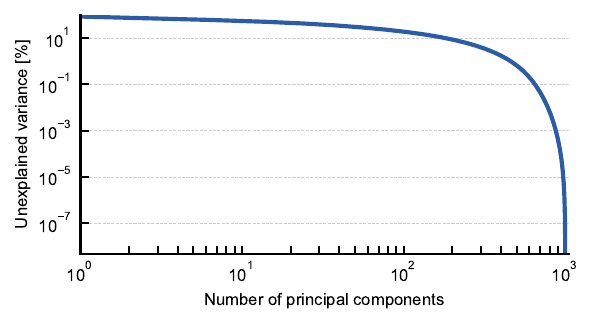}
    \caption{Residual variance as a function of the number of principal components. The steady decrease with more components shows a high intrinsic nature of the dimensionality of the dataset
    }
    \label{fig:pca-spectrum}
\end{figure}

\subsection{PET-MAD latent features}
For the high-dimensional description, we use 
the last-layer features of the trained PET-MAD model \cite{Mazitov2025}, that provide a 512-sized token describing each $i-$atom-centered environment in a given structure, $\boldsymbol{\xi}(A_i)$.
Given that we aim to characterize \emph{structures} rather than environments, we describe each configuration using a 1024-vector obtained by concatenating the entry-wise mean and standard deviation of the environment features
\begin{equation*}
\boldsymbol{\Xi}(A) = \left[\frac{1}{N_A} \sum_i \boldsymbol{\xi}(A_i),  \sqrt{\frac{1}{N_A} \sum_i (\boldsymbol{\xi}(A_i) -\left<\boldsymbol{\xi}(A_i)\right>)^2} \right].
\end{equation*}
This choice leads to an intensive description (if one replicates a periodic structure, the features don't change), but it is capable of describing the degree of inhomogeneity of a structure.

\begin{figure*}[tpb]
    \centering
    \includegraphics[width=\linewidth]{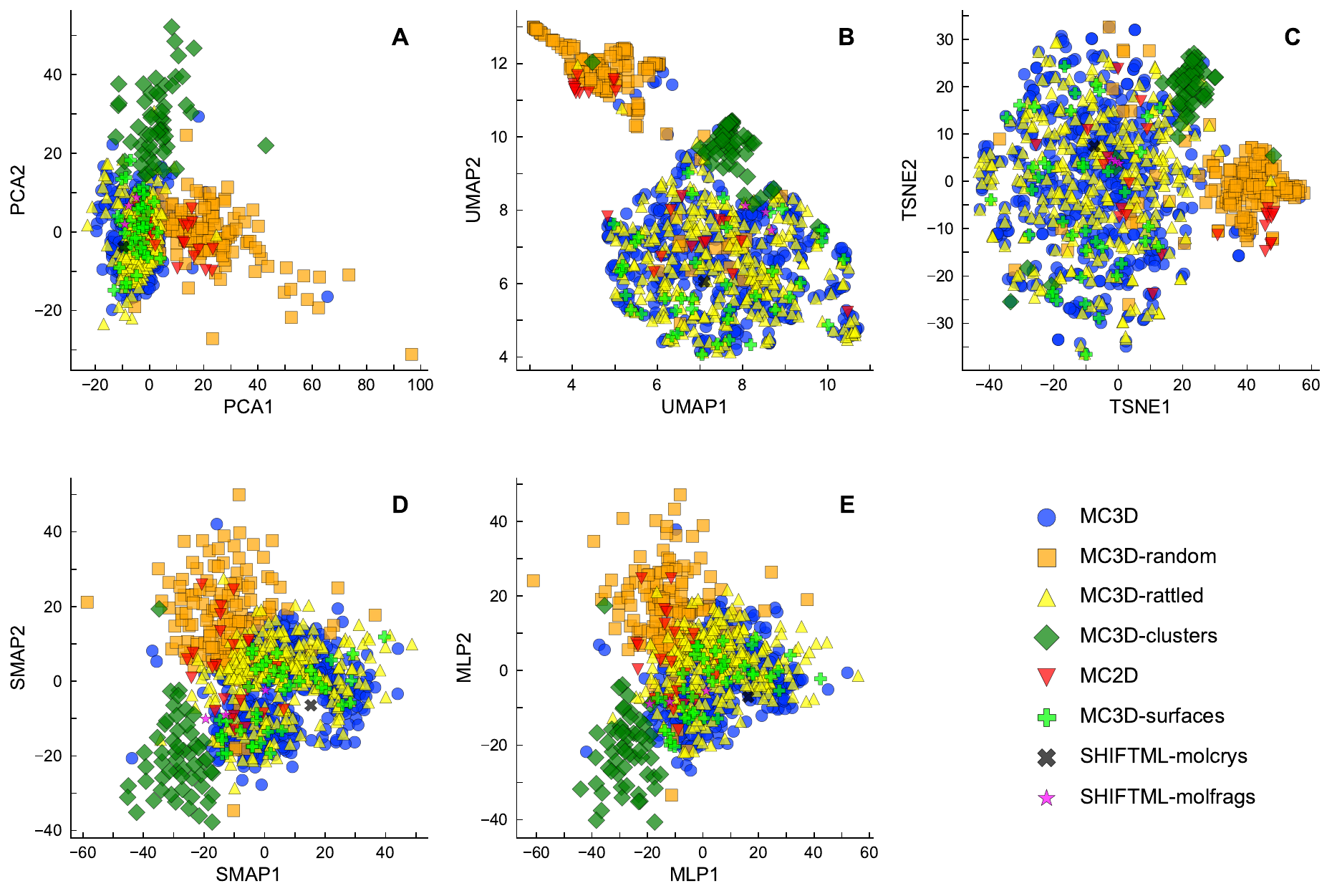}
    \caption{Two-dimensional projections of 1,000 representative MAD structures, selected via farthest-point sampling, based on high-dimensional features from the PET-MAD model. (a) PCA, (b) UMAP, (c) t-SNE, (d) Sketch-map projection, and (e) MLP-predicted sketch-map projection, learned to map high-dimensional descriptors to 2D space. The axes labels indicate the method used to determine the low-dimensional latent space in each map.}
\label{fig:sketchmap_landmarks}
\end{figure*}

A histogram of the Euclidean distances between pairs of configuration within each of the MAD subsets (Fig.~\ref{fig:dist-histograms}) demonstrates how the latent features capture the diversity of each subset, with the molecular datasets (that cover a small portion of chemical space) being peaked as small inter-configuration distance, and the more diverse MC3D-derived structures having a tail of large distances -- with the highest diversity corresponding to the MC3D-random subset.

\subsection{Dimensionality reduction}
Even though these histograms provide a way to qualitatively measure diversity in the high-dimensional feature space, an intuitive visualization requires performing a dimensionality reduction starting from the feature vectors. 
A principal component analysis (PCA) shows that the intrinsic dimensionality of the dataset is high, with a slow and smooth decay of the residual variance with the number of components included (Fig.~\ref{fig:pca-spectrum}). 
As a consequence, the low-dimensional representation is bound to be lossy, and to distort the relation between different structures. For this reason, we compare different several non-linear dimensionality-reduction algorithms (Figure \ref{fig:sketchmap_landmarks}).
We look for a projection that separates the different parts of PET-MAD in an intuitive manner, and that reflects the different degree of diversity of the various subsets as measured by the histograms of Euclidean distances between descriptor vectors (Fig.~\ref{fig:dist-histograms}).
For instance, we would expect the MC3D and MC3D-rattled to occupy roughly the same portion of space, and the MC3D-random structures to cover a much larger area, that overlaps in part with all the bulk structures; surfaces and low-dimensional structures should be at least partly separated from the bulk configurations; the organic molecules and crystals should be concentrated in a narrower region.

\begin{figure*}[tpb]
    \centering
    \includegraphics[width=\linewidth]{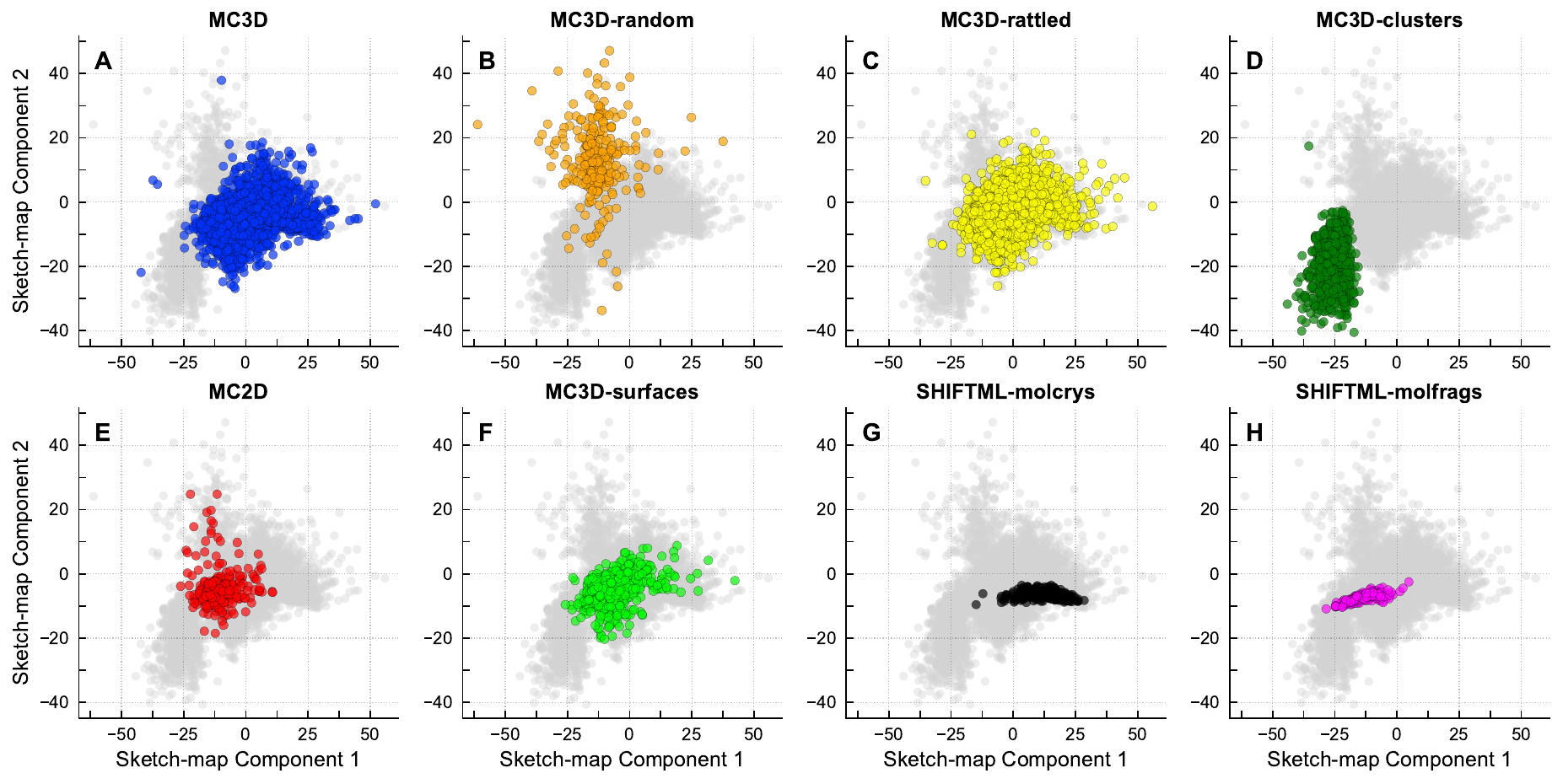}
    \caption{Two-dimensional sketch-map projection of subsets of MAD, built upon the last-layer features of the PET-MAD model}
    \label{fig:sketchmap_mad}
\end{figure*}

We extract 1000 landmark structures from the MAD dataset using a farthest point sampling strategy~\cite{imba+18jcp} using Euclidean distances between the high-dimensional feature vectors. By construction, these points serve as the most representative entries of the dataset and allow to analyse the overall coverage of the configuration space.
Their UMAP~\cite{umap} and t-SNE~\cite{tsne} projections (Fig.~\ref{fig:sketchmap_landmarks}b,c respectively) are only partly consistent with our list of requirements. Both concentrate the MC3D-random structures and the clusters in a narrow region, and mix completely the bulk MC3D structures and the surfaces. 
The MC3D space is fragmented into clusters that, upon inspection, are chemically homogeneous, even though there is no reason to expect that such clustering could be exaustive or meaningful (e.g. it would always be possible to create mixed structures that should interpolate between any pair of clusters). 
This tendency to ``over-cluster'' is a known issue with t-SNE~\cite{wattenberg2016how} and UMAP; in many ways, a simple PCA projection (Fig.~\ref{fig:sketchmap_landmarks}a) reflects more closely our requirements, with contiguous projections of the main classes of structures, and the extremely diverse MC3D-random structures covering a large portion of the map, that overlaps only partly with the bulk inorganic materials that have less outlandish compositions.
To incorporate non-linearity into the projection in a more controlled manner, we used sketch-map (SMAP)\cite{ceri+11pnas}, a method originally developed to analyze atomistic trajectories, that optimizes a multi-dimensional-scaling-like loss, transformed by sigmoid functions so that it aims to reproduce proximity, rather than Euclidean distance, between configurations in low and high dimension (see Section~\ref{sec:visualization} for a brief overview of the method). The resulting projection (Fig.~\ref{fig:sketchmap_landmarks}d) provides better separation of distinct subsets of MAD. 
Even though an out-of-sample projection can be performed to embed new data points on top of a sketch-map representation of landmarks, this is not very convenient -- as it involves an iterative optimization for each new point. For this reason, we train a multi-layer perceptron (MLP) to reproduce the embedding of the landmarks (Fig.~\ref{fig:sketchmap_landmarks}e), which matches nicely the landmark distribution from an explicit sketch-map optimization.

\begin{figure*}[tpb]
    \centering
    \includegraphics[width=\linewidth]{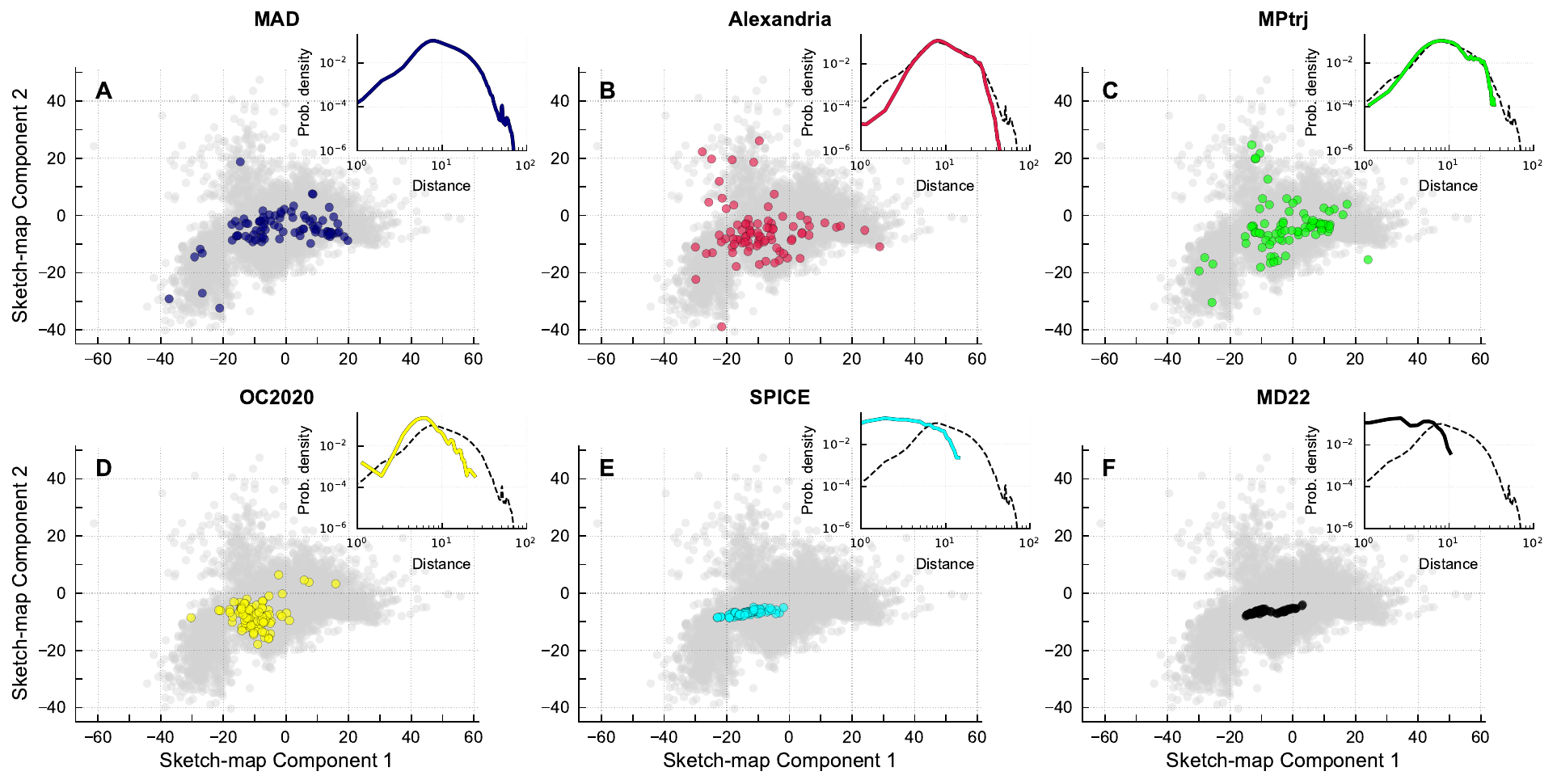}
    \caption{Two-dimentional projections of MAD dataset and popular benchmarks, using MLP-trained SMAP projections of PET-MAD last-layer embeddings. The grayscale points in the background correspond to the full test subset of the MAD dataset, and the colored points in each panel to the same, small set of 85 structures randomly selected from each dataset. Insets shows the histogram of the Euclidean distances between the highlighted structures in each panel, with the histogram of distances within the MAD dataset plotted for reference. 
    }
    \label{fig:sketchmap_mptraj}
\end{figure*}

We then use this MLP approximation of the sketch-map embedding to project all structures in the MAD test set onto the low-dimensional latent space.
The projection of individual subsets of MAD in this latent space (Figure~\ref{fig:sketchmap_mad}) corresponds to that of the landmarks, and is broadly consistent with our requirements, with MC3D ideal and distorted bulk structures overlapping almost perfectly, the randomized structures covering a broad (and partly overlapping) region, molecular bulk solids covering a narrower range, and structures of lower dimensionality being progressively shifted to the bottom-left. 

\subsection{Comparison with other datasets}

This dimensionality reduction can also be used to assess the MAD coverage relative to the existing benchmark datasets, by plotting them jointly on the same map using the same framework based on PET last-layer features and sketch-map-fitted dimensionality reduction.
Figure \ref{fig:sketchmap_mptraj} presents 2D projections of approximately selected structures from MAD and five datasets, Alexandria \cite{Schmidt2023, Wang2023}, MPtrj (MACE-MP-0 Val) \cite{MPtrj}, SPICE \cite{SPICE}, MD22 \cite{MD22}, and OC2020 (S2EF) \cite{OC20}.
The SPICE and MD17 dataset are  narrowly focused in the center of the map, roughly in the area where the SHIFTML subsets are projected. 
Their low chemical diversity is apparent also in the histogram of feature-space distances, peaked at short values. 
OC2020 contains molecules adsorbed on surfaces, providing data to study heterogeneous catalysis, and is projected roughly in the region associated with MC3D surface subsets. 
In this highly-compressed projection, MPtrj and Alexandria appear to cover roughly a similar space as MAD, which however has a considerably larger spread in the third dimension (cf. the interactive visualization provided in the SI), which is also evident in the shorter tails of the distance histograms.

\section{Methods}

\subsection{Details of the dataset construction}\label{sec:dataset_details}

In the case of MC3D, MC2D, SHIFTML-molcrys, and SHIFTML-molfrags subsets of MAD, we used previously published structures and recomputed them using a consistent set of DFT settings. 
For other subsets, we initialized the structure generation protocol with random MC3D crystals and performed different transformations to increase the overall coverage of the configuration space.
Rattled structures in the MC3D-rattled subset were obtained by selecting a random MC3D crystal with more than one atom in the unit cell and applying a Gaussian noise to the Cartesian coordinates of each atom with a zero mean and a standard deviation equal to 20\% of the corresponding covalent radii. 
The MC3D-random subset was built by assigning a random set of atom types to the lattice sites of a randomly chosen MC3D crystal, followed by an isotropic adjustment of the cell volume to a total atomic volume computed based on the covalent radii of the included elements. 
For the MC3D-surface subset, we created the surface slabs of randomly chosen MC3D crystals by cleaving them along a randomly chosen symmetrically distinct crystallographic plane with a maximum value of the Miller index $(hkl)$ equal to 3 and assuring orthogonality of the normal lattice vector to a surface plane. 
Finally, the structures in the MC3D-cluster subset were created by cutting a random atomic environment of 2 to 8 atoms of a randomly chosen atom from a random MC3D crystal. 
We discard structures for which the DFT calculations did not converge, and a few outliers with forces exceeding a very large threshold (100 eV/Å\ for MC3D-rattled and MC3D-random and 10 eV/Å\ for the other subsets). Details of the first principles calculations are provided below in Section \ref{sec:dft_details}. 
We generate a random 80:10:10 train:validation:test split of the dataset, which we recommend using to benchmark ML potentials.

To facilitate a consistent comparison with models trained against datasets (e.g. Alexandria\cite{Schmidt2023, Wang2023}, MPtraj\cite{MPtrj} or Matbench\cite{WBM-dataset, Matbench} that are computed using VASP and a PBE functional, using Hubbard U corrections\cite{jain+13aplm} for transition metals oxides), we also create a MAD-benchmark dataset that is computed both with MAD and MPtraj-like settings.
For the MAD dataset, 50 structures were randomly sampled from each test subset and recalculated using MPtrj DFT parameters. Non-converged and outlier structures were excluded, resulting in a final set of 322 structures. Similarly, the OC2020 benchmark subset includes 89 structures, constructed by sampling 100 structures and removing non-converged cases. 
The Alexandria benchmark consists of 150 structures, incorporating randomly selected 50 samples from Alexandria-2D and Alexandria-3D-gopt. 
For the MD22 benchmark, 25 structures were randomly selected from each MD22's subset (Ac-Ala3-NHMe, AT-AT, DHA, Stachyose, AT-AT-CG-CG, Buckyball-Catcher, double-walled-nanotube). The SPICE benchmark subset consists of 100 randomly chosen neutral molecules. Finally, MPtrj (MACE-MP-0 validation subset) was reduced to 153 structures with the exclusion of 1D wire structures.
For all these datasets, structures for which either type of DFT calculations did not converge were removed.

\subsection{Details of the electronic-structure reference}\label{sec:dft_details}

To maintain a consistent level of theory across the MAD dataset, all calculations were intentionally conducted without spin polarization. 
This choice introduces obvious errors in the description of strongly magnetic materials, avoids the likely convergence to inconsistent magnetization states, and mitigates issues related to the incorrect magnetic descriptions for elements with strong electronic correlations within spin-polarized density functional theory.
The calculations were performed with Quantum Espresso v7.2 \cite{gian+09jpcm} compiled with the SIRIUS libraries \cite{SIRIUS}. The workflows were managed by the AiiDA framework \cite{Pizzi2016,Huber2020,Uhrin2021}.
We used the PBEsol functional \cite{perdew2008pbesol}, which is designed to have better accuracy than its very similar PBE counterpart for inorganic solids, even though both are not very accurate for several classes of materials (e.g. molecular compounds). 
Once again, we prioritize stability and consistency for highly diverse systems over the accuracy against experiments. 
The behavior of semi-core electrons and their interaction with valence electrons was described using the standard solid-state pseudopotentials library (SSSP) v1.2 (efficiency set) \cite{Prandini2018} selecting plane-wave and charge-density cutoffs (110 Ry and 1320 Ry) corresponding to the largest recommended values across the 85 elements we considered.
Convergence for metallic systems was facilitated using a smearing of the electronic occupations, using the Marzari-Vanderbilt-DeVita-Payne cold smearing function\cite{marz+99prl} with a spread of 0.01 Ry. The Brillouin zone was sampled with a $\Gamma$-centered grid resolution of 0.125 Å$^{-1}$, in periodic dimensions, while non-periodic dimensions were treated with a single k-point. 
To prevent interaction through periodic boundary conditions in non-periodic structures, we applied the Sohier-Calandra-Mauri method \cite{moun+18nn} for 2D systems and the Martyna-Tuckerman correction \cite{Martyna1999Reciprocal} for 0D systems, with a 25~Å vacuum along  non-periodic directions to ensure convergence. 
Additionally, a compositional baseline based on isolated atom energies was subtracted from the DFT energies to improve the numerical stability during model training.
These DFT settings achieved a convergence rate exceeding $95\%$ for most of the MAD subsets, described in Section \ref{sec:dataset_details}, with the exception of the MC3D-random structures. Due to the completely arbitrary combination of elements, these configurations had a much lower convergence rate of approximately $55\%$.

\subsection{Details of the dataset visualization technique}\label{sec:visualization}

To explore the structural and chemical diversity and coverage of the MAD dataset and simplify the comparison with other atomistic datasets, we used sketch-map, a non-linear dimensionality reduction algorithm \cite{sketchmap} designed as an extension to multi-dimensional scaling~\cite{mds}.
The idea of is to project high-dimensional data into a low-dimensional space while preserving \emph{proximity} rather than the Euclidean distances between high-dimensional and low-dimensional vectors ($D_{ij}$ and $d_{ij}$ respectively).

More specifically, sketch-map minimizes a stress function 
\begin{equation}
    \mathcal{L}_{\text{sm}} = \sum_{i \neq j} w_{ij} (F(D_{ij}) - f(d_{ij}))^2,
    \label{eq:transformed_loss}
\end{equation}
where \(D_{ij}\) and \(d_{ij}\) are the Euclidean distances between pairs of points \(i\) and \(j\) in high- and low-dimensional spaces, respectively, and \(w_{ij}\) are weights that can be included, e.g. as the product of the number of structures within the Voronoi cell of each reference landmark point.
\(F\) and \(f\) are sigmoid functions which determine the classification of  ``far'' and  ``near'' pairs:
\begin{equation*}
    F(D_{ij}) = 1 - \left[ 1 + \left( 2^{A/B} - 1 \right) \left( \frac{D_{ij}}{\sigma} \right)^A \right]^{-B/A}
\end{equation*}
and
\begin{equation*}
   f(d_{ij}) = 1 - \left[1 + \left( 2^{a/b} - 1 \right) \left( \frac{d_{ij}}{\sigma} \right) ^a \right] ^{-b/a}.
\end{equation*}
The parameter \(\sigma\) controls the distance scale at which the sigmoid functions switches from \(0\) to \(1\). The parameters \(A\), \(B\), \(a\), and \(b\) define the steepness and asymptotic behaviour of the sigmoid transitions at short and large distances, and can be used to adjust the sensitivity of the notion of proximity in high and low dimension.

Given the computational complexity of applying sketch-map directly to large datasets, we first selected 1,000 landmark structures from the MAD test set using farthest-point sampling, a method that iteratively chooses structures to maximize coverage of configuration space~\cite{imba+18jcp}. 
The sketch-map projection was then performed for the landmarks with a sigmoid transformation defined by parameters: $\sigma=7$, $A=4$, and $B=2$ for the high-dimensional space; and $a=2$, $b=2$ for the low-dimensional space, following the hyperparameter selection methodology described in \cite{smap_method}. %
The landmark projections are initialized to a 2D PCA, followed by a sequence of local and global optimization steps. Finally, an iterative optimization is performed including a third low-dimensional component to allow for more descriptive 3D representation.
After having obtained a 3D projection of the landmarks, we train a simple neural network to reproduce the sketch-map embedding in a simpler, and less computationally demanding, way. 
We use an simple Multi-Layer Perceptron (MLP) architecture, with three hidden layers with ReLU activation functions, to map high-dimensional PET-MAD descriptors to 3D sketch-map coordinates. 
We trained the model on the landmarks, using an 80:20 train-validation split and using SmoothL1Loss\cite{girshick2015fastrcnn} to assess the error in the projection. 
The MLP was then applied to project the remaining points from the MAD validation split and the benchmark datasets.

To visualize the resulting projections and analyze their compositions, we used Chemiscope \cite{frau+20joss,chemiscope}, a visualization tool that allows to interactively explore atomistic structures and their properties. It enables to inspect the low-dimensional projections and associated structures and their properties. Figure \ref{code:archetypal} shows a code snippet demonstrating how to use Chemiscope to visualise a new dataset with the PET-MAD model.

\begin{figure}[h]
\centering
\noindent\begin{minipage}{0.5\textwidth}
\begin{lstlisting}
import ase.io

import chemiscope
from pet_mad.explore import PETMADFeaturizer

# Read structures
frames = ase.io.read("dataset.xyz", ":")

# Generate visualisation of the dataset
chemiscope.explore(frames, 
   featurize=PETMADFeaturizer(
       version="latest"
   )
)
\end{lstlisting}
\end{minipage}
\caption{Example of Python code for visualization a dataset using Chemiscope with PET-MAD descriptors mapped to sketch-map coordinates.}
\label{code:archetypal}
\end{figure}

Here, \texttt{frames} represents a list of ASE\cite{ase} compatible input structures, and \texttt{featurizer} extracts the PET-MAD descriptors, which are then mapped to the sketch-map coordinates using the trained MLP.

\section{Data record}

The dataset is made available as a record~\cite{matcloud25a} within the Materials Cloud~\cite{Materials_Cloud} Archive, which is a FAIR repository dedicated to materials-science simulations.
The data is stored in the format of AiiDA archive files, that contain the full provenance graph of the calculations and can be accessed using the tools provided by the AiiDA package~\cite{pizz+16cms}. 
In addition to this monolithic database, we extract parts of the data in a more compact, and easier-to-access format (the extended XYZ format, that stores energies, stresses and lattice parameters in the header, and atom types, positions and forces as space-separated entries).
Cartesian coordinates, energies, forces, and stresses are given in \AA{}, eV, eV/\AA{}, and eV/\AA$^3$, respectively.

More specifically, we provide:
\begin{itemize}
\item The MAD dataset, with an 80:10:10 train:validation:test split as used in training the PET-MAD model: \emph{mad-train.xyz}, \emph{mad-val.xyz}, \emph{mad-test.xyz}. The subsets are indicated as a field in the header.
\item AiiDA database archives for each subset of the MAD dataset: \emph{mad-mc3d.aiida}, \emph{mad-mc3d-rattled.aiida},  \emph{mad-mc3d-random.aiida}, 
\emph{mad-mc3d-clusters.aiida},  \emph{mad-mc3d-surfaces.aiida},  \emph{mad-mc2d.aiida},  \emph{mad-shiftml-molcrys.aiida},  \emph{mad-shiftml-molfrags.aiida}.
\item The MAD benchmark dataset, containing a selection of MAD test, MPtrj, Alexandria, SPICE, MD22 and OC2020 datasets, computed with both MAD DFT settings, and MPtrj DFT settings. These are provided as two separate files: \emph{mad-bench-mad-settings.xyz}, \emph{mad-bench-mptrj-settings.xyz}. The parent dataset is indicated as a field in the header.
\item A zipped folder \emph{mad-aiida-aux.zip} with AiiDA database archives for the MAD benchmark calculations, as well as some auxiliary calculations done for the MAD dataset.
\end{itemize}

To facilitate visualization, we also provide chemiscope visualization files corresponding to Figure \ref{fig:sketchmap_landmarks}, Figure \ref{fig:sketchmap_mad}, and Figure \ref{fig:sketchmap_mptraj} -- containing 2D or 3D latent space projections, as well as energies for each structure. These files, named \texttt{mad-landmarks.chemiscope.json.gz},  \texttt{mad-subsets.chemiscope.json.gz}, and  \texttt{mad-bench.chemiscope.json.gz}, respectively, can be viewed and interacted using the Chemiscope web interface at \url{http://chemiscope.org} (or directly via custom links from the Materials Cloud Archive record page) or programmatically via the Chemiscope API using \texttt{chemiscope.show\_input("mad-
subsets.chemiscope.json.gz")} in a Jupyter notebook.

More details on how to interact with the data are provided in the data record page \cite{matcloud25a}.

\begin{acknowledgments}
AM and MC acknowledge support from an Industrial Grant from BASF SE, and Dr. Olivier Enger from BASF SE personally. 
MB, GF, GP and MC acknowledge funding by the NCCR MARVEL, a National Centre of Competence in Research, funded by the Swiss National Science Foundation (grant number 205602).
SC in particular was supported by an Inspire Potentials Fellowship.
GF acknowledges support from the Platform for Advanced Scientific Computing (PASC).

We acknowledge the EuroHPC Joint Undertaking for awarding this project access to the EuroHPC supercomputer LUMI, hosted by CSC (Finland) and the LUMI consortium through a EuroHPC Extreme Scale Access call (project EHPC-EXT-2022E01-048). Additionally, we acknowledge access to Piz Daint, Eiger and Daint Alps at the Swiss National Supercomputing Centre, Switzerland with the project IDs s1219, s1243, s1287 on Piz Daint, project ID mr31 under MARVEL's share on Eiger, and project ID lp26 on Daint Alps.

\end{acknowledgments}

\end{document}